\documentclass[twoside,fleqn]{article}
\usepackage{espcrc2}
\usepackage{graphicx}
\newcommand{\be}{\begin{equation}}
\newcommand{\ee}{\end{equation}}

\newcommand{\numberofconfigs}{222 }

\title{ 
\vspace{-2.8cm}
\hfill \rm \null \hfill
\hbox{\normalsize ADP-03-130/T566} \\
\hfill \hbox{\normalsize DESY-03-209} \\
\vspace{1.65cm}
Hybrid and Exotic Mesons from FLIC Fermions}

\author{J.N.~Hedditch\address[CSSM]{Special Research Center for the
    Subatomic Structure of Matter, and		\\
    Department of Physics, University of Adelaide
    Adelaide SA 5005  Australia}
    D.~B.~Leinweber\addressmark[CSSM],
    A.~G.~Williams\addressmark[CSSM]\thanks{Presented by A.~G.~Williams
    at Lattice '03} 
    and
    J.~M.~Zanotti\addressmark[CSSM]\address[DESY]{John von
    Neumann-Institut f\"ur Computing NIC, \\
    Deutches Elektronen-Synchrotron DESY, D-15738 Zeuthen, Germany}
} 
 
\begin{document}

\begin{abstract}
The spectral properties of hybrid meson interpolating fields are
investigated.  The quantum numbers of the meson are carried by
smeared-source fermion operators and highly-improved chromo-electric
and -magnetic field operators composed with APE-smeared links.  The
effective masses of standard and hybrid operators indicate that the
ground state meson is effectively isolated using both standard and
hybrid interpolating fields.  Focus is placed on interpolating fields
in which the large spinor components of the quark and antiquark
fields are merged.  In particular, the effective mass of the exotic
$1^{-+}$ meson is reported. Further, we report some values for excited
mesonic states using a variational process.
\end{abstract}

\maketitle

\section{INTRODUCTION}

Major experimental efforts are currently aimed at determining the
possible existence of exotic mesons; mesons having quantum numbers
that cannot be carried by the minimal Fock space component of a
quark-antiquark pair.  Of particular mention is the proposed program
of the GlueX collaboration associated with the forthcoming upgrade of
the Jefferson Laboratory facility.  The observation of exotic states
and the determination of their properties would elucidate aspects of
QCD which are relatively unexplored.

The quantum numbers $J^{PC} = 0^{+-},\ 0^{--},\ 1^{-+},$ etc.  cannot
be carried by a quark-antiquark pair in a ground-state $S$-wave.
Lattice QCD calculations exploring the non-trivial role of explicit
gluon degrees of freedom in carrying the quantum numbers of the meson
suggest that exotic meson states do indeed exist and have a mass the
order of 2~GeV \cite{importantPapers}.  These findings are further
supported here.  

\section{SIMULATION METHODOLOGY}

Exotic quantum numbers can be constructed by merging standard local
interpolating fields $\overline{q}^a(x) \Gamma q^a(x)$ with
chromo-electric, $E_i^{ab}(x)$, or chromo-magnetic fields,
$B_i^{ab}(x)$.  The $J^{PC}$ quantum numbers of the interpolator are
derived from the direct product of those associated with the quark
bilinear and $E_i^{ab}\ (1^{--})$ or $B_i^{ab}\ (1^{+-})$.
For example, combining the vector current of the $\rho$ meson with a
chromo-magnetic field, $1^{--} \otimes 1^{+-}$ provides $0^{-+} \oplus
1^{-+} \oplus 2^{-+}$ with the $0^{-+}$: $\bar q^a \gamma_i q^b
B_i^{ab}$ ($\pi$ meson) and the $1^{-+}$: $\epsilon_{ijk} \bar q^a
\gamma_i q^b B_j^{ab}$ ({\bf exotic}).
We restrict ourselves to the lowest energy-dimension operators,
as these provide better signal with smaller statistical errors.
Table \ref{interpolators} summarizes the standard and hybrid
interpolating fields explored herein.  

\begin{table*}
\caption{$J^{PC}$ quantum numbers and their associated meson
  interpolating fields.}
\label{interpolators}
\begin{tabular}{cccc}
\hline
\noalign{\smallskip}
$0^{++}$ & $0^{+-}$ & $0^{-+}$ & $0^{--}$ \\
\hline
\noalign{\smallskip}
$\bar{q}^a q^a$ & $\bar{q}^a \gamma_4 q^a$ &  $\bar{q}^a \gamma_5 q^a$ & $-i \bar{q}^a \gamma_5 \gamma_j E^{ab}_j q^b$ \\
$- i \bar{q}^a \gamma_j E^{ab}_j q^b$ & $ \bar{q}^a \gamma_5 \gamma_j B^{ab}_j q^b$ & $\bar{q}^a \gamma_5 \gamma_4 q^a$ \\
$-  \bar{q}^a \gamma_j \gamma_4 \gamma_5 B^{ab}_j q^b$ & & $- \bar{q}^a \gamma_j B^{ab}_j q^b$ \\
$-  \bar{q}^a \gamma_j \gamma_4 E^{ab}_j q^b$ & & $- \bar{q}^a \gamma_4 \gamma_j B^{ab}_j q^b$        \\
\noalign{\smallskip}
\hline
\noalign{\medskip}
\hline
\noalign{\smallskip}
$1^{++}$ & $1^{+-}$ & $1^{-+}$ & $1^{--}$ \\
\hline
\noalign{\smallskip}
  $- i \bar{q}^a \gamma_5 \gamma_j q^a$
& $- i \bar{q}^a \gamma_5 \gamma_4 \gamma_j q^a$ 
& $\bar{q}^a \gamma_4 E^{ab}_j q^b$
& $- i \bar{q}^a \gamma_j q^a$\\
  $ i \bar{q}^a \gamma_4 B^{ab}_j q^b$
& $i \bar{q}^a B^{ab}_j q^b $ 
& $- \epsilon_{jkl} \bar{q}^a \gamma_k B^{ab}_l q^b$
& $\bar{q}^a E^{ab}_j q^b$\\
  $ i \epsilon_{jkl} \bar{q}^a \gamma_k E^{ab}_l q^b$
& $\bar{q}^a \gamma_5 E^{ab}_j q^b $
& $\epsilon_{jkl} \bar{q}^a \gamma_4 \gamma_k B^{ab}_l q^b$
& $ - i \bar{q}^a \gamma_5 B^{ab}_j q^b $\\
   $ i \epsilon_{jkl} \bar{q}^a \gamma_k \gamma_4 E^{ab}_l q^b$
& $\bar{q}^a \gamma_5 \gamma_4 E^{ab}_j q^b$
& $- i \epsilon_{jkl} \bar{q}^a \gamma_5 \gamma_4 \gamma_k E^{ab}_l q^b$
& $i \bar{q}^a \gamma_4 \gamma_5 B^{ab}_j q^b$\\
\noalign{\smallskip}
\hline
\end{tabular}
\end{table*}

The formulation of effective interpolating fields for the creation and
annihilation of exotic meson states continues to be an active area of
research. Here we consider local interpolating fields. Gauge-invariant Gaussian
smearing \cite{Gusken:qx,Zanotti:2003fx} is applied at the fermion
source ($t=3$), and local sinks are used to maintain strong signal in
the two-point correlation functions.  Our experience is that
smeared-smeared correlation functions suffer an increase
in error bar size relative to smeared-local correlation functions.
Chromo-electric and -magnetic fields are created from APE-smeared
links \cite{ape} at both the source and sink using the highly-improved
${\mathcal O}(a^4)$-improved lattice field strength tensor
\cite{Bilson-Thompson:2002jk}.  The smearing fraction $\alpha = 0.7$
(keeping 0.3 of the original link).\cite{Bonnet:2000dc}.

Propagators are generated using the fat-link irrelevant clover (FLIC)
fermion action \cite{FATJAMES} where the irrelevant Wilson and clover
operators of the fermion action are constructed using fat links while
the relevant operators use the untouched (thin) gauge links.  FLIC
fermions provide a new form of nonperturbative ${\cal O}(a)$
improvement \cite{Leinweber:2002bw,inPrep} where near-continuum
results are obtained at finite lattice spacing.  Access to the light
quark mass regime is enabled by the improved chiral properties of the
lattice fermion action \cite{inPrep}.

The following results are based on \numberofconfigs mean-field ${\cal
O}(a^2)$-improved Luscher-Weisz \cite{Luscher:1984xn} gauge fields on
a $16^3 \times 32$ lattice at $\beta = 4.60$ providing a lattice
spacing of $a = 0.122(2)$ fm set by the string tension $\sqrt{\sigma}
= 440$ MeV.

\section{RESULTS}
\label{results}

\begin{figure}[tb]
\begin{center}
{\includegraphics[height=\hsize,angle=90]{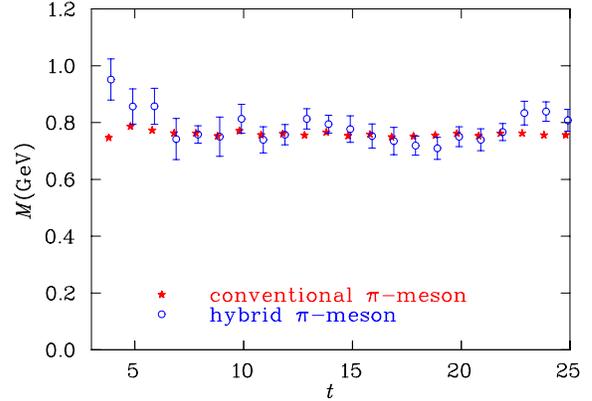}}
\vspace*{-1.4cm}
\caption{Effective mass plot for correlation functions of the standard
axial-vector pion interpolator $\bar{q}^a \gamma_5 \gamma_4 q^a$ and
the hybrid pion interpolator $\bar{q}^a \gamma_4 \gamma_j B^{ab}_j
q^b$.}
\label{pion22pion44}
\end{center}
\vspace*{-1.0cm}
\end{figure}

\begin{table*}[t]
\caption{Meson masses as a function of the hopping parameter $\kappa$.}
\label{masses_conventional}
\begin{tabular}{rcccccc}
\hline
\noalign{\smallskip}
   &  & \multicolumn{5}{c}{Mass(GeV)} \\
$J^{PC}$&Operator&$\kappa=0.1260$&$\kappa=0.1266$ &
$\kappa=0.1273$ & $\kappa=0.1279$ & $\kappa=0.1286$ 
\\ 
\hline 
\noalign{\smallskip}
$\pi:\, 0^{-+}$  & $\bar{q}^a \gamma_5 q^a$ &
$0.937\pm .005 $ &
$0.861 \pm .006$ &
$0.764 \pm .006$ &
$0.672 \pm .007$ &
$0.545\pm .007$ 
\\

        & $\bar{q}^a \gamma_5 \gamma_4 q^a$ & 
$0.932\pm .006 $&
$0.856 \pm .007$&
$0.761 \pm .007$&
$0.670 \pm .007$ &
$0.548 \pm .007$ 
\\

        & $- \bar{q}^a \gamma_j B^{ab}_j q^b$ & 
$0.999\pm .062$&
$0.918 \pm .066$&
$0.816 \pm .072$&
$0.722 \pm .076$ &
$0.561 \pm .081$
\\

       & $- \bar{q}^a \gamma_4 \gamma_j B^{ab}_j q^b$ & 
$0.964 \pm .045 $&
$0.888 \pm .045$&
$0.791 \pm .045$&
$0.696 \pm .045$ &
$0.561 \pm .046$
\\

\noalign{\smallskip}
$b_1:\, 1^{+-}$  &  $- i \bar{q}^a \gamma_5 \gamma_4 \gamma_j q^a$ &
$1.675\pm        .012$ &
$1.633\pm        .013$ &
$1.585\pm        .013$ &
$1.548\pm        .015$ &
$1.513\pm        .017$
\\

        &  $i \bar{q}^a B^{ab}_j q^b $&
$2.373\pm        .256$ &
$2.319\pm        .246$ &
$2.264\pm        .233$ &
$2.225\pm        .222$ &
$2.223\pm        .224$
\\
\noalign{\smallskip}
$\rho:\, 1^{--}$  &  $-i \bar{q}^a \gamma_j q^a $&
$1.184\pm        .007$ &
$1.131\pm        .008$ &
$1.070\pm        .009$ &
$1.016\pm        .010$ &
$0.954\pm        .013$
\\

        &  $-i \bar{q}^a \gamma_5 B^{ab}_j q^b $&
$1.255\pm        .138$ &
$1.178\pm       .117$ &
$1.077\pm       .086$ &
$0.977\pm       .138$ &
$0.836\pm       .174$
\\

       &  $i \bar{q}^a \gamma_4 \gamma_5 B^{ab}_j q^b $&
$1.244\pm       .086$ &
$1.172\pm       .087$ &
$1.079\pm       .088$ &
$0.987\pm       .092$ &
$0.861\pm       .102$
\\
\noalign{\smallskip}
$1^{-+}$ &$- \epsilon_{jkl} \bar{q}^a \gamma_k B^{ab}_l q^b$ &
$2.892 \pm .475$ &
$2.881 \pm .475$ &
$2.884 \pm .479$ &
$2.908 \pm .500$ &
$2.969 \pm .516$ 
\\
\noalign{\smallskip}
\hline
\end{tabular}
\end{table*}

\begin{figure}[tb]
\begin{center}
{\includegraphics[height=\hsize,angle=90]{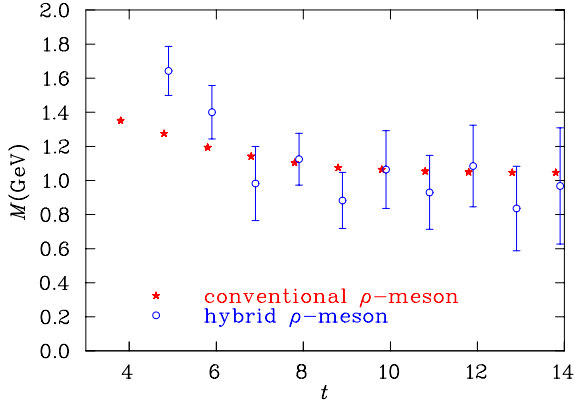}}
\vspace*{-1.4cm}
\caption{Effective mass plot for correlation functions of the standard
$\rho$-meson interpolator $\bar{q}^a \gamma_j q^a $ and
the hybrid $\rho$ interpolator $\bar{q}^a \gamma_4 \gamma_5
B^{ab}_j q^b $. }
\label{rho11rho33}
\end{center}
\vspace*{-1.0cm}
\end{figure}

\begin{figure}[tb]
\begin{center}
{\includegraphics[height=\hsize,angle=90]{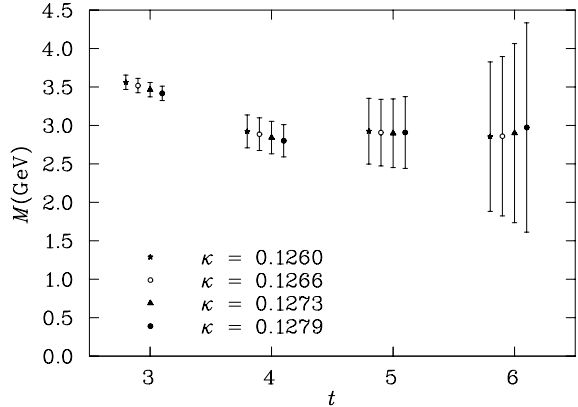}}
\vspace*{-1.4cm}
\caption{Effective mass plot of the $1^{-+}$ exotic meson obtained
from the hybrid interpolating field $\epsilon_{jkl} \bar{q}^a \gamma_k
B^{ab}_l q^b$.  }
\label{1mp22mass}
\end{center}
\vspace*{-1.0cm}
\end{figure}

Of the hybrid interpolators listed in Table \ref{interpolators}, only
the interpolating fields merging the large spinor components of the
quark and antiquark fields provide a clear mass plateau.  Figure
\ref{pion22pion44} illustrates the effective masses $M(t) = - \log
(G(t+1)/G(t))$, obtained from the second and fourth pion ($0^{-+}$)
interpolators of Table \ref{interpolators} for our intermediate quark
mass ($m_\pi^2 \sim 0.6\ {\rm GeV}^2$).  Here $G(t)$ is the standard
two-point function projected to zero three-momentum.  Excellent
agreement is seen between the standard and hybrid interpolator-based
correlation functions.  Similar results are seen in Fig.\
\ref{rho11rho33} comparing effective masses obtained from the first
and third $\rho$-meson ($1^{--}$) interpolators of Table
\ref{interpolators}.  The effective mass plot for the exotic $1^{-+}$
meson is illustrated in Fig.\ \ref{1mp22mass}, where a plateau at
early times is observed confirming the existence of the exotic
$1^{-+}$.

Table \ref{masses_conventional} summarizes these preliminary results.
Further work on this topic will focus on extrapolating results to
physical quark masses and using variational techniques to extract the
masses of excited mesonic states.
\vspace*{0.3cm} This research is supported by the Australian National
Computing Facility for Lattice Gauge Theory and the Australian
Research Council.


\end{document}